\documentclass
[aps,prl,twocolumn,a4paper,superscriptaddress,amsmath,amssymb]{revtex4}%
\usepackage[dvips]{graphicx}
\usepackage[dvips]{color}
\usepackage{bbold}
\usepackage{amsmath}
\usepackage{amsfonts}
\usepackage{amssymb}%
\providecommand{\U}[1]{\protect\rule{.1in}{.1in}}
\begin{document}
\title{Scattering Theory of Gilbert Damping}
\author{Arne Brataas}
\email{Arne.Brataas@ntnu.no}
\affiliation{Department of Physics, Norwegian University of Science and Technology, N-7491
Trondheim, Norway}
\author{Yaroslav Tserkovnyak}
\affiliation{Department of Physics and Astronomy, University of California, Los Angeles,
California 90095, USA}
\author{Gerrit E. W. Bauer}
\affiliation{Kavli Institute of NanoScience, Delft University of Technology, Lorentzweg 1,
2628 CJ Delft, The Netherlands}

\begin{abstract}
The magnetization dynamics of a single domain ferromagnet in contact with a
thermal bath is studied by scattering theory. We recover the
Landau-Liftshitz-Gilbert equation and express the effective fields and Gilbert
damping tensor in terms of the scattering matrix. Dissipation of magnetic
energy equals energy current pumped out of the system by the
time-dependent magnetization, with separable spin-relaxation induced bulk and 
spin-pumping generated interface contributions. In linear response, our scattering theory
for the Gilbert damping tensor is equivalent with the Kubo formalism.
\end{abstract}


\maketitle


Magnetization relaxation is a collective many-body phenomenon that remains
intriguing despite decades of theoretical and experimental investigations. 
It is important in topics of current interest since it
determines the magnetization dynamics and noise in magnetic memory devices and 
state-of-the-art magnetoelectronic experiments on current-induced magnetization 
dynamics \cite{Stiles:top06}. Magnetization relaxation is often described in 
terms of a damping torque in
the phenomenological Landau-Lifshitz-Gilbert (LLG) equation%
\begin{equation}
\frac{1}{\gamma}\frac{d \mathbf{M}}{d \tau}=-\mathbf{M}%
\times\mathbf{H}_{\mathrm{eff}}+\mathbf{M}\times\left[  \frac{\tilde
{G}(\mathbf{M})}{\gamma^{2}M_{s}^{2}}\frac{d \mathbf{M}}{d \tau
}\right]  \text{,}\label{LLG}%
\end{equation}
where $\mathbf{M}$ is the magnetization vector, $\gamma=g\mu_{B}/\hbar$ is the
gyromagnetic ratio in terms of the $g$ factor and the Bohr magneton $\mu_{B}$,
and $M_{s}=|\mathbf{M}|$ is the saturation magnetization. Usually, the Gilbert
damping $\tilde{G}(\mathbf{M})$ is assumed to be a scalar and isotropic
parameter, but in general it is a symmetric $3\times3$ tensor. The LLG
equation has been derived microscopically \cite{Heinrich:pss67} and successfully describes the
measured response of ferromagnetic bulk materials and thin films in terms of a
few material-specific parameters that are accessible to
ferromagnetic-resonance (FMR) experiments \cite{Bland:book05}. We focus in the
following on small ferromagnets in which the spatial degrees of freedom are
frozen out (macrospin model). Gilbert damping predicts a stricly linear 
dependence of FMR linewidts on frequency. This distinguishes it from 
inhomogenous broadening associated with dephasing of the global precession, 
which typically induces a weaker frequency dependence as well as a 
zero-frequency contribution.

\begin{figure}
  \includegraphics[scale=0.666]{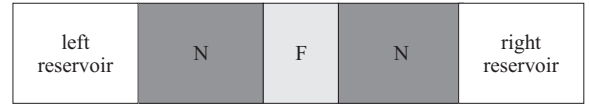}
  \caption{\label{fig:scattering} Schematic picture of a ferromagnet (F) in contact with 
a thermal bath via metallic normal metal leads (N).}
\end{figure}

The effective magnetic field $\mathbf{H}_{\mathrm{eff}}=-\partial
F/\partial\mathbf{M}$ is the derivative of the free energy $F$ of the magnetic
system in an external magnetic field $\mathbf{H}_{\mathrm{ext}}$, including
the classical magnetic dipolar field $\mathbf{H}_{d}$. When the ferromagnet is
part of an open system as in Fig.\ \ref{fig:scattering}, $-\partial F/\partial\mathbf{M}$ can be expressed in
terms of a scattering S-matrix, quite analogous to the interlayer
exchange coupling between ferromagnetic layers \cite{Bruno:prb95}. The
scattering matrix is defined in the space of the transport channels that
connect a scattering region (the sample) to thermodynamic (left and right) reservoirs by
electric contacts that are modeled by ideal leads. Scattering matrices also
contain information to describe giant magnetoresistance, spin pumping
and spin battery, and current-induced magnetization dynamics in layered
normal-metal (N)$\mid$ferromagnet (F) systems
\cite{Bruno:prb95,Waintal:prb00,Tserkovnyak:prl02}.

In the following we demonstrate that scattering theory can be also used to
compute the Gilbert damping tensor $\tilde{G}(\mathbf{M}).$ The energy loss rate
of the scattering region can be described in terms of the time-dependent S-matrix.
Here, we generalize the theory of adiabatic quantum pumping to describe dissipation 
in a metallic ferromagnet. Our idea is to evaluate the energy pumping out of the ferromagnet
and to relate this to the energy loss of the LLG equation.  We find that the Gilbert
phenomenology is valid beyond the linear response regime of small
magnetization amplitudes. The only approximation that is necessary to derive
Eq.~(\ref{LLG}) including $\tilde{G}(\mathbf{M})$ is the (adiabatic)
assumption that the frequency $\omega$ of the magnetization dynamics is slow
compared to the relevant internal energy scales set by the exchange splitting
$\Delta$. The LLG phenomenology works so well because $\hbar\omega\ll\Delta$
safely holds for most ferromagnets.

Gilbert damping in transition-metal ferromagnets is generally believed to stem
from spin-orbit interaction in combination with impurity scattering that
transfers magnetic energy to itinerant quasiparticles \cite{Bland:book05}.  The subsequent
drainage of the energy out of the electronic system, \textit{e.g.} by
inelastic scattering via phonons, is believed to be a fast process that does
not limit the overall damping. Our key assumption is adiabaticiy, meaning that the precession 
frequency goes to zero before letting the sample size become large. The magnetization dynamics 
then heats up the entire magnetic system by a tiny amount that escapes via the contacts. 
The leakage heat current then equals the total dissipation rate. For sufficiently large 
samples, bulk heat production is insensitive to the
contact details and can be identified as an additive contribution to the total
heat current that escapes via the contacts. The chemical potential is set by
the reservoirs, which means that (in the absence of an intentional bias) the
sample is then always very close to equilibrium. The S-matrix expanded to linear order in the magnetization dynamics and the Kubo linear response formalisms 
should give identical results, which we will explicitly demonstrate. The role of the
infinitesimal inelastic scattering that guarantees causality in the Kubo
approach is in the scattering approach taken over by the coupling to the reservoirs. Since the electron-phonon relaxation is not expected to directly impede the overall rate of magnetic energy dissipation, we do not need to explicitly include it in our treatment. The energy flow supported by the leads, thus, appears in our model to be carried entirely by electrons irrespective of whether the energy is actually carried by phonons, in case the electrons relax by inelastic scattering before reaching the leads. So we are able to compute the magnetization damping, but not, e.g., how the sample heats up by it .

According to Eq.~(\ref{LLG}), the time derivative of the energy reads%
\begin{equation}
\dot{E}=\mathbf{H}_{\mathrm{eff}}\mathbf{\cdot}d \mathbf{M/}d
\tau=(1/\gamma^{2})\mathbf{\dot{m}}\left[  \tilde{G}(\mathbf{m})\mathbf{\dot
{m}}\right]  \mathbf{,}\label{Energy_rate_LLG}%
\end{equation}
in terms of the magnetization direction unit vector $\mathbf{m}=\mathbf{M}%
/M_{s}$ and $\mathbf{\dot{m}}=d \mathbf{m} / d \tau$.  We now
develop the scattering theory for a ferromagnet connected to two reservoirs by
normal metal leads as shown in Fig.\ \ref{fig:scattering}. The total energy pumping into both leads $I_{E}%
^{\mathrm{(pump)}}$ at low temperatures reads
\cite{Avron:prl01,Moskalets:prb02}%
\begin{equation}
I_{E}^{\mathrm{(pump)}}=(\hbar/4\pi)\mathrm{Tr}\dot{S}\dot{S}^{\dag
},\label{Energy_pumping}%
\end{equation}
where $\dot{S}=dS/d\tau$ and $S$ is the S-matrix at the Fermi energy:
\begin{equation}
S(\mathbf{m})=\left(
\begin{array}
[c]{cc}%
r & t^{\prime}\\
t & r^{\prime}%
\end{array}
\right)  \,.
\end{equation}
$r$ and $t$ ($r^{\prime}$ and $t^{\prime}$) are the reflection and
transmission matrices spanned by the transport channels and spin states for an
incoming wave from the left (right). The generalization to finite temperatures
is possible but requires knowledge of the energy dependence of the S-matrix
around the Fermi energy \cite{Moskalets:prb02}. The S-matrix changes
parametrically with the time-dependent variation of the magnetization
$S(\tau)=S(\mathbf{m}(\tau))$. We obtain the Gilbert damping tensor in terms
of the S-matrix by equating the energy pumping by the magnetic system
(\ref{Energy_pumping}) with the energy loss expression (\ref{Energy_rate_LLG}%
), $\dot{E}=I_{E}^{\text{(pump)}}$. Consequently
\begin{equation}
G_{ij}(\mathbf{m})=\frac{\gamma^{2}\hbar}{4\pi}\operatorname{Re}\left\{
\mathrm{Tr}\left[  \frac{\partial S}{\partial m_{i}}\frac{\partial S^{\dag}%
}{\partial m_{j}}\right]  \right\}  \,,\label{Gilbert_scattering}%
\end{equation}
which is our main result.

The remainder of our paper serves three purposes. We show that (i) the
S-matrix formalism expanded to linear response is equivalent to Kubo linear response formalism,
demonstrate that (ii) energy pumping reduces to interface spin pumping in the
absence of spin relaxation in the scattering region, and (iii) use a simple
2-band toy model with spin-flip scattering to explicitly show that we can
identify both the disorder and interface (spin-pumping) magnetization damping as
additive contributions to the Gilbert damping.

Analogous to the Fisher-Lee relation between Kubo conductivity and the
Landauer formula \cite{Fisher:prb81} we will now prove that the Gilbert
damping in terms of S-matrix (\ref{Gilbert_scattering}) is consistent with the
conventional derivation of the magnetization damping by the linear response
formalism.  To this end we chose a
generic mean-field Hamiltonian that depends on the magnetization direction
$\mathbf{m}$: $\hat{H}=\hat{H}(\mathbf{m})$ describes the system in 
Fig.\ \ref{fig:scattering}. $\hat{H}$ can describe realistic
band structures as computed by density-functional theory including exchange-correlation effects and spin-orbit coupling as well normal and spin-orbit induced scattering off impurities. The energy dissipation is $\dot{E}=\langle
d\hat{H}/d\tau\rangle$, where $\langle\dots\rangle$ denotes the expectation value
for the non-equilibrium state. In linear response, we expand the magnetization
direction $\mathbf{m}(t)$ around the equilibrium magnetization direction
$\mathbf{m}_{0}$,
\begin{equation}
\mathbf{m}(\tau)\mathbf{=m}_{0}+\mathbf{u}(\tau).\label{magn_linear}%
\end{equation}
The Hamiltonian can be linearized as $\hat{H}=\hat{H}_{\mathrm{st}}+u_{i}%
(\tau)\partial_{i}\hat{H}$, where $\hat{H}_{\mathrm{st}}\equiv\hat
{H}(\mathbf{m}_{0})$ is the static Hamiltonian and $\partial_{i}\hat{H}%
\equiv\partial_{u_{i}}\hat{H}(\mathbf{m}_{0})$, where summation over repeated
indices $i=x,y,z$ is implied. To lowest order $\dot{E}=\dot{u}_{i}%
(\tau)\langle\partial_{i}\hat{H}\rangle$, where
\begin{equation}
\langle\partial_{i}\hat{H}\rangle=\langle\partial_{i}\hat{H}\rangle_{0}%
+\int_{-\infty}^{\infty}d\tau^{\prime}\chi_{ij}(\tau-\tau^{\prime})u_{j}%
(\tau^{\prime})\,.\label{eff_field_lres}%
\end{equation}
$\langle\dots\rangle_{0}$ denotes equilibrium expectation value and the
retarded correlation function is
\begin{equation}
\chi_{ij}(\tau-\tau^{\prime})=-\frac{i}{\hbar}\theta(\tau-\tau^{\prime
})\left\langle [\partial_{i}\hat{H}(\tau),\partial_{j}\hat{H}(\tau^{\prime
})]\right\rangle _{0}\,\label{corr}%
\end{equation}
in the interaction picture for the time evolution. In order to arrive at the
adiabatic (Gilbert) damping the magnetization dynamics has to be sufficiently
slow such that $u_{j}(\tau)\approx u_{j}(t)+\left(  \tau-t\right)  \dot{u}%
_{j}(t)$. Since $\mathbf{m}^{2}=1$ and hence $\mathbf{\dot{m}\cdot m}=0$
\cite{Simanek:prb03}
\begin{equation}
\dot{E}=i\partial_{\omega}\chi_{ij}(\omega\rightarrow0)\dot{u}_{i}\dot{u}%
_{j},\label{E_diss_lres_ad}%
\end{equation}
where $\chi_{ij}(\omega)=\int_{-\infty}^{\infty}d\tau\chi_{ij}(\tau
)\exp(i\omega\tau)$. \ Next, we use the scattering states as the basis for
expressing the correlation function (\ref{corr}). The Hamiltonian consists of
a free-electron part and a scattering potential: $\hat{H}=\hat{H}_{0}+\hat
{V}(\mathbf{m})$. We denote the unperturbed eigenstates of the free-electron
Hamiltonian $\hat{H}_{0}=-\hbar^{2}\nabla^{2}/2m$ at energy $\epsilon$ by
$|\varphi_{s,q}(\epsilon)\rangle$, where $s=l,r$ denotes propagation direction
and $q$ transverse quantum number. The potential $\hat{V}(\mathbf{m})$
scatters the particles between these free-electron states. The outgoing (+)
and incoming wave (-) eigenstates $|\psi_{s,q}^{(\pm)}(\epsilon)\rangle$ of
the static Hamiltonian $\hat{H}_{\mathrm{st}}$ fulfill the completeness
conditions $\langle\psi_{s,q}^{(\pm)}(\epsilon)|\psi_{s^{\prime},q^{\prime}%
}^{(\pm)}(\epsilon^{\prime})\rangle=\delta_{s,s^{\prime}}\delta_{q,q^{\prime}%
}\delta(\epsilon-\epsilon^{\prime})$ \cite{MelloKumar:book04}. These wave
functions can be expressed as $|\psi_{s}^{(\pm)}(\epsilon)\rangle=[1+\hat
{G}_{\mathrm{st}}^{(\pm)}\hat{V}_{\mathrm{st}}]|\varphi_{s}(\epsilon)\rangle$,
where the static retarded (+) and advanced (-) Green functions are $\hat
{G}_{\mathrm{st}}^{(\pm)}(\epsilon)=(\epsilon\pm i\eta-\hat{H}_{\mathrm{st}%
})^{-1}$ and $\eta$ is a positive infinitesimal. By expanding $\chi
_{ij}(\omega)$ in the basis of the outgoing wave functions $|\psi_{s}%
^{(+)}\rangle$, the low-temperature linear response leads to the following
energy dissipation (\ref{E_diss_lres_ad}) in the adiabatic limit
\begin{equation}
\dot{E}=-\pi \hbar \dot{u}_{i}\dot{u}_{j}\left\langle \psi_{s,q}^{(+)}|\partial
_{i}\hat{H}|\psi_{s^{\prime},q^{\prime}}^{(+)}\right\rangle \left\langle
\psi_{s^{\prime},q^{\prime}}^{(+)}|\partial_{j}\hat{H}|\psi_{s,q}%
^{(+)}\right\rangle \,,\label{E rate Kubo}%
\end{equation}
with wave functions evaluated at the Fermi energy $\epsilon_{F}$. 

In order to compare the linear response result, Eq. (\ref{E rate Kubo}), with that
of the scattering theory, Eq. (\ref{Gilbert_scattering}), we introduce the T-matrix
$\hat{T}$ as $\hat{S}(\epsilon;\mathbf{m})=1-2\pi i\hat{T}(\epsilon
;\mathbf{m})$, where $\hat{T}=\hat{V}[1+\hat{G}^{(+)}\hat{T}]$ in terms of the
full Green function $\hat{G}^{(+)}(\epsilon,\mathbf{m})=[\epsilon+i\eta
-\hat{H}(\mathbf{m})]^{-1}$.
Although the adiabatic energy pumping (\ref{Gilbert_scattering}) is valid for any magnitude of slow magnetization dynamics, in order to make connection to the linear-response formalism we should consider small magnetization changes to the equilibrium values as described by Eq.~(\ref{magn_linear}). We then find
\begin{equation}
\partial_{\tau}\hat{T}=\left[  1+\hat{V}_{\mathrm{st}}\hat{G}_{\mathrm{st}%
}^{(+)}\right]  \dot{u}_{i}\partial_{i}\hat{H}\left[  1+\hat{G}_{\mathrm{st}%
}^{(+)}\hat{V}_{\mathrm{st}}\right]  \,.
\end{equation}
into Eq.~(\ref{Gilbert_scattering}) and using the completeness of the scattering
states, we recover Eq. (\ref{E rate Kubo}).

Our S-matrix approach generalizes the theory of (nonlocal) spin pumping and
enhanced Gilbert damping in thin ferromagnets \cite{Tserkovnyak:prl02}: by
conservation of the total angular momentum the  spin current pumped into the
surrounding conductors implies an additional damping torque that enhances the
bulk Gilbert damping. Spin pumping is an N$\mid$F interfacial effect that
becomes important in thin ferromagnetic films \cite{Heinrich:prl03}. In the
absence of spin relaxation in the scattering region, the S-matrix can be
decomposed as $S(\mathbf{m})=S_{\uparrow}(1+\hat{\boldsymbol{\sigma}}%
\cdot\mathbf{m})/2+S_{\downarrow}(1-\hat{\boldsymbol{\sigma}}\cdot
\mathbf{m})/2$, where
$\hat{\boldsymbol{\sigma}}$ is a vector of Pauli matrices. In this case,
$\mathrm{Tr}$\thinspace$\left(  \partial_{\tau}S\right)  \left(
\partial_{\tau}S\right)  ^{\dagger}=A_{r}\mathbf{\dot{m}}^{2}$, where
$A_{r}=\mathrm{Tr}[1-\operatorname{Re}S_{\uparrow}S_{\downarrow}^{\dag}]$ and
the trace is over the orbital degrees of freedom only. We recover the diagonal
and isotropic Gilbert damping tensor: $G_{ij}=\delta_{ij}G$ derived earlier
\cite{Tserkovnyak:prl02}, where
\begin{equation}
G=\gamma M_{s}\alpha=\frac{\left(  g\mu_{B}\right)  ^{2}}{4\pi\hbar}A_{r}\,.
\end{equation}

Finally, we illustrate by a model calculation that we can obtain magnetization
damping by both spin-relaxation and interface spin-pumping from the S-matrix. We
consider a thin film ferromagnet in the two-band Stoner model embedded in a free-electron metal
\begin{equation}
\hat{H}=-\frac{\hbar^{2}}{2m}\nabla^{2}+\delta(x)\hat{V}(\boldsymbol{\rho})\,,
\end{equation}
where the in-plane coordinate of the ferromagnet is $\boldsymbol{\rho}$ \ and
the normal coordinate is $x.$ The spin-dependent potential $\hat
{V}(\boldsymbol{\rho})$ consists of the mean-field exchange interaction
oriented along the magnetization direction $\mathbf{m}$ and magnetic disorder
in the form of magnetic impurities $\mathbf{S}_{i}$ %

\begin{equation}
\hat{V}(\boldsymbol{\rho})=\nu\hat{\boldsymbol{\sigma}}\cdot\mathbf{m}%
+\sum_{i}\zeta_{i}\hat{\boldsymbol{\sigma}}\cdot\mathbf{S}_{i}\delta
(\boldsymbol{\rho}-\boldsymbol{\rho}_{i}),
\end{equation}
which are randomly oriented and distributed in the film at $x=0$. Impurities in combination with spin-orbit 
coupling will give similar contributions as magnetic impurities to Gilbert damping. Our derivation of the S-matrix closely follows
Ref.~\cite{Brataas:prb94}. The 2-component spinor wave function can be written
as $\Psi(x,\boldsymbol{\rho})=\sum_{\mathbf{k}_{\Vert}}c_{\mathbf{k}_{\Vert}%
}(x)\Phi_{\mathbf{k}_{\Vert}}(\boldsymbol{\rho})$, where the transverse wave
function is $\Phi_{\mathbf{k}_{\Vert}}(\boldsymbol{\rho})=\exp(i\mathbf{k}%
_{\Vert}\cdot\boldsymbol{\rho})/\sqrt{A}$ for the cross-sectional area $A$.
The effective one-dimensional equation for the longitudinal part of the wave
function is then%
\begin{equation}
\left[  \frac{d^{2}}{dx^{2}}+k_{\bot}^{2}\right]  c_{\mathbf{k}_{\Vert}%
}(x)=\sum_{\mathbf{k}_{\Vert}^{\prime}}\tilde{\Gamma}_{\mathbf{k}_{\Vert
},\mathbf{k}_{\Vert}^{\prime}}c_{\mathbf{k}_{\Vert}}(0)\delta(x)\,,
\end{equation}
where the matrix elements are defined by $\tilde{\Gamma}_{\mathbf{k}_{\Vert
},\mathbf{k}_{\Vert}^{\prime}}=(2m/\hbar^{2})\int d\boldsymbol{\rho}%
\,\Phi_{\mathbf{k}_{\Vert}}^{\ast}(\boldsymbol{\rho})\hat{V}(\boldsymbol{\rho
})\Phi_{\mathbf{k}_{\Vert}^{\prime}}(\boldsymbol{\rho)}$ and the longitudinal
wave vector $k_{\bot}$ is defined by $k_{\bot}^{2}=2m\epsilon_{F}/\hbar
^{2}-\mathbf{k}_{\Vert}^{2}$.
For an incoming electron from the left, the longitudinal wave function is
\begin{equation}
c_{\mathbf{k}_{\Vert}s}=\frac{\chi_{s}}{\sqrt{k_{\bot}}}\left\{
\begin{array}
[c]{cc}%
e^{ik_{\bot}x}\delta_{\mathbf{k}_{\Vert}s,\mathbf{k}_{\Vert}^{\prime}%
s^{\prime}}+e^{-ik_{\bot}x}r_{\mathbf{k}_{\Vert}s,\mathbf{k}_{\Vert}^{\prime
}s^{\prime}} & ,x<0\\
e^{ik_{\bot}x}t_{\mathbf{k}_{\Vert}s,\mathbf{k}_{\Vert}^{\prime}s^{\prime}} &
,x>0
\end{array}
\right.  \,,
\end{equation}
where $s=\uparrow,\downarrow$ and $\chi_{\uparrow}=(1,0)^{\dag}$ and
$\chi_{\downarrow}=(0,1)^{\dag}$. Inversion symmetry dictates that $t^{\prime
}=t$ and $r=r^{\prime}$. Continuity of the wave function requires $1+r=t$. The
energy pumping (\ref{Energy_pumping}) then simplifies to $I_{E}^{\text{(pump)}%
}=\hbar\mathrm{Tr}\left(  \dot{t}\dot{t}^{\dag}\right)  /\pi$. Flux continuity
gives $t=(1+i\hat{\Gamma})^{-1}$, where $\hat{\Gamma}_{\mathbf{k}_{\Vert
}s,\mathbf{k}_{\Vert}^{\prime}s^{\prime}}=\chi_{s}^{\dag}\hat{\Gamma
}_{\mathbf{k}_{\Vert}s,\mathbf{k}_{\Vert}^{\prime}s^{\prime}}\chi_{s^{\prime}%
}\left(  4k_{\bot}k_{\bot}\right)  ^{-1/2}$.

In the absence of spin-flip scattering, the transmission coefficient is
diagonal in the transverse momentum: $t_{\mathbf{k}_{\Vert}}^{(0)}%
=[1-i\eta_{\bot}\mathbf{\sigma}\cdot\mathbf{m}]/(1+\eta_{\bot}^{2})$, where
$\eta_{\bot}=m\nu/(\hbar^{2}k_{\bot})$. The nonlocal (spin-pumping) Gilbert
damping is then isotropic, $G_{ij}(\mathbf{m})=\delta_{ij}G^{\prime}$,
\begin{equation}
G^{\prime}=\frac{2\nu^{2}\hbar}{\pi}\sum_{\mathbf{k}_{\Vert}}\frac{\eta_{\bot
}^{2}}{(1+\eta_{\bot}^{2})^{2}}\,.\label{G_nonloc}%
\end{equation}
It can be shown that $G^{\prime}$ is a function of the ratio between the
exchange splitting versus the Fermi wave vector, $\eta_{F}=m\nu/(\hbar
^{2}k_{F})$. $G^{\prime}$ vanishes in the limits $\eta_{F}\ll1$ (nonmagnetic
systems) and $\eta_{F}\gg1$ (strong ferromagnet).

We include weak spin-flip scattering by expanding the transmission coefficient
$t$ to second order in the spin-orbit interaction, $t\approx\left[
1+t_{0}i\hat{\Gamma}_{\mathrm{sf}}-\left(  t_{0}i\hat{\Gamma}_{\mathrm{sf}%
}\right)  ^{2}\right]  t_{0}$, which inserted into
Eq.~(\ref{Gilbert_scattering}) leads to an in general anisotropic Gilbert
damping. Ensemble averaging over all random spin configurations and positions
after considerable but straightforward algebra leads to the isotropic result
$G_{ij}(\mathbf{m})=\delta_{ij}G$
\begin{equation}
G=G^{(\text{\textrm{int}})}+G^{\prime}\,\label{G}%
\end{equation}
where $G^{\prime}$ is defined in Eq.~(\ref{G_nonloc}). The \textquotedblleft
bulk\textquotedblright\ contribution to the damping is caused by the
spin-relaxation due to the magnetic disorder
\begin{equation}
G^{(\text{\textrm{int}})}=N_{s}S^{2}\zeta^{2}\xi\,,\label{G_intrinsic}%
\end{equation}
where $N_{s}$ is the number of magnetic impurities, $S$ is the impurity spin,
$\zeta$ is the average strength of the magnetic impurity scattering, and
$\xi=\xi(\eta_{F})$ is a complicated expression that vanishes when
$\eta_{F}$ is either very small or very large. Eq. (\ref{G}) proves that
Eq. (\ref{Gilbert_scattering}) incorporates the \textquotedblleft
bulk\textquotedblright\ contribution to the Gilbert damping, which grows with
the number of spin-flip scatterers, in addition to interface damping. We could
have derived $G^{(\text{\textrm{int}})}$ [Eq.~(\ref{G_intrinsic})] as well by
the Kubo formula for the Gilbert damping.

The Gilbert damping has been computed before based on the Kubo formalism based
on first-principles electronic band structures \cite{Gilmore:prl07}. However,
the \textit{ab initio} appeal is somewhat reduced by additional approximations
such as the relaxation time approximation and the neglect of disorder vertex
corrections. An advantage of the scattering theory of Gilbert damping is its
suitability for modern \textit{ab initio} techniques of spin transport that do
not suffer from these drawbacks \cite{Zwierzycki:pstat:08}. When extended to include 
spin-orbit coupling and magnetic disorder the Gilbert damping can be obtained without 
additional costs according to Eq.~(\ref{Gilbert_scattering}). Bulk and interface contributions
can be readily separated by inspection of the sample thickness dependence of
the Gilbert damping.

Phonons are important for the understanding of damping at elevated temperatures, which we
do not explicitly discuss. They can be included by a temperature-dependent relaxation time \cite{Gilmore:prl07} or, in our case, structural disorder. A microscopic treatment of phonon excitations requires extension of the formalism to inelastic scattering, which is beyond the scope of the present paper.

In conclusion, we hope that our alternative formalism of Gilbert damping will
stimulate \textit{ab initio} electronic structure calculations as a function
of material and disorder. By comparison with FMR studies on thin ferromagnetic
films this should lead to a better understanding of dissipation in magnetic systems.

\begin{acknowledgments}
This work was supported in part by the Research Council of Norway, Grants
Nos.~158518/143 and 158547/431, and EC Contract IST-033749 ``DynaMax."
\end{acknowledgments}


\end{document}